\documentclass[11pt]{article}

\usepackage{amsmath,amssymb,amscd,amsfonts,graphicx}
\allowdisplaybreaks[4]

\begin{document}

\begin{flushright} 

\end{flushright} 

\vspace{0.1cm}

\begin{center}
  {\LARGE
Numerical approach to SUSY quantum mechanics 

and 

the gauge/gravity duality
  }
\end{center}
\vspace{0.1cm}
\vspace{0.1cm}
 \begin{center}

          Masanori H{\sc anada}\footnote{
Affiliation since October 2010: 
Department of Physics, University of Washington, 
 Seattle, WA 98195-1560, USA; 
 mhanada@u.washington.edu
} 
\\
\vspace{5mm}
Department of Particle Physics and Astrophysics\\
Weizmann Institute of Science\\
Rehovot 76100, Israel   

 \end{center}
\vspace{1.5cm}

\begin{center}
  {\bf Abstract}
\end{center}

We demonstrate that Monte-Carlo simulation is a practical tool 
to study nonperturbative aspects of supersymmetric quantum mechanics. 
As an example we study D0-brane quantum mechanics 
in the context of superstring theory. 
Numerical data nicely reproduce predictions from gravity side, 
including the coupling constant dependence of the string $\alpha'$ correction. 
This strongly suggests the duality to hold beyond the supergravity approximation.  
Although detail of the stringy correction cannot be obtained 
by state-of-the-art techniques in gravity side, 
in the matrix quantum mechanics we can obtain concrete values. 
Therefore the Monte-Carlo simulation combined with the duality 
provides a powerful tool to study the superstring theory.

\newpage


\section{Introduction}
Supersymmetric Yang-Mills (SYM) theories play prominent roles in theoretical particle physics. 
Especially, maximally supersymmetric theories are of crucial importance 
for superstring/M theory \cite{BFSS,IKKT,MatrixString,Maldacena:1997re,Itzhaki:1998dd}. 
Given that most interesting questions can be answered only through nonperturbative study, 
it is important to construct theoretical frameworks for that. 
However, it is not a straightforward task because of the notorious 
difficulties of lattice supersymmetry (SUSY). 

Why is lattice SUSY difficult? In lattice gauge theory, gauge symmetry is 
kept unbroken. Therefore gauge symmetry breaking terms are not generated  
by radiative corrections. On the other hand, supersymmetry cannot be preserved 
completely, because the SUSY algebra contains infinitesimal translation, 
which is broken on lattice by construction. Therefore, even if a given 
lattice theory converges to a supersymmetric theory at tree level, 
SUSY breaking operators can be generated radiatively. In order to control the divergence  
one needs some exact symmetries at discretized level. In 4d ${\cal N}=1$ pure SYM, by keeping 
the chiral symmetry one can obtain the correct supersymmetric continuum limit 
\cite{Kaplan:1983sk}\footnote{
For recent numerical studies, see \cite{Giedt:2008xm}. 
}. 
In several extended SYM in less than four dimensions, by keeping a part 
of supersymmetries intact, SUSY breaking operators are forbidden at least 
to all order in perturbation theory \cite{Kaplan:2002wv}. 
(A similar statement holds for the Wess-Zumino model, and numerically tested in 
\cite{Catterall:2001fr}.)  

In one dimension (i.e. supersymmetric matrix quantum mechanics), 
the situation is much easier. Because the theory is UV finite, 
one does not have to rely on exact symmetries and hence 
a simple momentum cutoff prescription works \cite{Hanada:2007ti}. 
In fact, as demonstrated in \cite{Hanada:2007ti}, the momentum cutoff method is 
much more powerful than a usual lattice regularization -- 
the convergence to the continuum and restoration of the supersymmetry become 
much faster, and the Fourier acceleration, which 
reduces the critical slowing down, can be implemented without any additional cost. 
As a result it is possible to perform detailed Monte-Carlo study. In this talk 
we show the result of the Monte-Carlo simulation of maximally supersymmetric 
matrix quantum mechanics, which describes a system of multiple D0-branes,   
and compare it with dual string theory. 
Numerical data nicely reproduce not only predictions from supergravity 
but also the string $\alpha'$ correction. 
This strongly suggests the duality to hold beyond the supergravity approximation.  
In fact, in the gravity side, the stringy correction cannot be evaluated completely  
because of the lack of the knowledge on the higher derivative correction to 
the supergravity action.  
In the matrix quantum mechanics, however, we can obtain concrete values numerically. 
Therefore the Monte-Carlo simulation provides a new powerful tool to study the superstring theory.  

Although we concentrate on 
a specific model relevant for the gauge/gravity duality, 
the same method applies to other theories as well.
Our message is {\it Monte-Carlo simulation is a powerful and practical 
tool to study (SUSY-) quantum mechanics}.

This talk is organized as follows. 
In \S~\ref{sec:duality} we briefly review the gauge/gravity duality. 
In \S~\ref{sec:simulation_setup} we explain the simulation method. Then in 
\S~\ref{sec:results} we provide the simulation results and their interpretations. 
The materials treated here appeared previously in 
\cite{Nishimura:2009xm}. In order to avoid the repetition, 
we put emphasis on an explanation of the duality, so that the physical meaning 
of the simulation results become clearer.    
\section{The gauge/gravity duality in the D0-brane system}\label{sec:duality}
The gauge/gravity duality conjecture \cite{Maldacena:1997re,Itzhaki:1998dd} claims  
{\it type II string theory on black $p$-brane background and 
maximally supersymmetric super Yang-Mills theory in $p+1$ dimensions are equivalent}. 
When $p=3$ this is known as the $AdS_5/CFT_4$ correspondence. 
In this paper we concentrate on $p=0$ case. The ``derivation" of the correspondence 
is the same as $p=3$ -- both SYM and (weakly-coupled) superstring theory descriptions are valid  
in a certain parameter region of a coincident D$p$-brane system, 
and hence they should be equivalent.  
$U(N)$ super Yang-Mills is obtained in the near horizon  
limit of the $N$-coincident D$p$-brane system, where bulk degrees of freedom decouple. 
In the same limit the black $p$-brane solution in type II supergravity 
reduces to the near extremal solution, and at large-$N$ and strong 't Hooft coupling 
the supergravity approximation is valid. 
The near extremal solution is given by 
\cite{Gibbons:1987ps,Horowitz:1991cd}\cite{Itzhaki:1998dd}
\begin{eqnarray}
ds^2
&=&
\alpha'\Biggl\{
\frac{U^{(7-p)/2}}{g_{YM}\sqrt{d_p N}}
\left[
-\left(
1-\frac{U_0^{7-p}}{U^{7-p}}
\right)dt^2
+
dy_{\parallel}^2
\right]
\nonumber\\
& &
\qquad
+
\frac{g_{YM}\sqrt{d_p N}}{U^{(7-p)/2}\left(1-\frac{U_0^{7-p}}{U^{7-p}}\right)}dU^2
+
g_{YM}\sqrt{d_p N}U^{(p-3)/2}d\Omega_{8-p}^2
\Biggl\},
\label{near extremal metric} 
\end{eqnarray}
\begin{eqnarray}
e^{\phi}=(2\pi)^{2-p}g_{YM}^2
\left(
\frac{d_p g_{YM}^2N}{U^{7-p}}
\right)^{\frac{3-p}{4}}
\end{eqnarray}
where $U$ is radial coordinate perpendicular to the black brane, 
$y_{\parallel}$ is 
the coordinate parallel to the brane, $\Omega_{8-p}$ represents 
the spherical coordinate of the transverse directions 
and 
a constant $d_p$ is given by 
$d_p=2^{7-2p}\pi^{(9-3p)/2}\Gamma((7-p)/2)$.  
$U_0$ represents the horizon of the black brane, 
which is related to the Hawking temperature $T$ as 
\begin{eqnarray}
T
=
\frac{(7-p)U_0^{(5-p)/2}}{4\pi\sqrt{d_p\lambda}}. 
\end{eqnarray}
Note that the gauge coupling constant $g_{YM}$ has dimension of $({\rm mass})^{(3-p)/2}$.   
As a result, an effective coupling constant describing the black hole thermodynamics is 
a dimensionless combination $T^{-(3-p)}\lambda$, 
where $\lambda=g_{YM}^2N$ is the 't Hooft coupling. 
For this reason, when $p<3$ ``strong coupling" is equivalent to ``low temperature".  
The dilaton $\phi$, and hence the string coupling $g_s=e^\phi$, depends on the radial coordinate, 
except for $p=3$. 
In order for the supergravity approximation to be valid, 
the metric should be weakly curved so that $\alpha'$ correction is negligible, 
and $g_s$ must be small so that the closed string loop correction is small. 
Roughly speaking, the latter corresponds to the planar limit ($N\to\infty$ with 
fixed 't Hooft coupling) and the former is achieved at strong 't Hooft coupling 
(low temperature). 

On the contrary to the superstring, SYM is defined nonperturbatively. 
Therefore, if the gauge/gravity duality is correct, it provides 
a nonperturbative formulation of superstring theory. 
As we will see, this viewpoint is important not only conceptually, but also {\it practically} -- 
once one puts SYM on computer, stringy correction to the supergravity, which is difficult to 
evaluate from the string theory, can be calculated numerically. We demonstrate it in \S~\ref{sec:results}. 

As an example of the correspondence, let us consider 
the ADM mass of the black brane per unit volume and the energy density of the gauge theory. 
They are predicted to be the same value, by identifying the Hawking temperature 
of the black brane to the temperature in the gauge theory.  
For $p=0$, the prediction from the supergravity is 
\begin{eqnarray}
\frac{E\lambda^{-1/3}}{N^2}
=
7.4 (T\lambda^{-1/3})^{14/5}, 
\end{eqnarray}
where the 't Hooft coupling $\lambda=g_{YM}^2N$ has dimension of $({\rm mass})^3$, 
$T$ is the temperature of the system and $E$ is the energy density. 
This relation is expected to hold at $N^{-10/21}\ll T\lambda^{-1/3}\ll 1$, 
with which the supergravity approximation is valid around the horizon. 
Note that the black 0-brane has a positive specific heat, in sharp contrast 
to the Schwarzschild black hole. 

When one applies $AdS/CFT$ duality to solve strongly coupled field theory, 
the Gubser-Klebanov-Polyakov-Witten (GKPW) relation \cite{GKPW}, 
which relates the anomalous dimension of the operators in CFT to 
mass of the supergravity modes, is very useful. 
Similar correspondence is proposed for $p\neq 3$ case as well. 
The basic idea of the GKPW relation is as follows. 
Let us consider a system of $N$-coincident D-branes, which gives 
black $p$-brane background, and one probe D-brane 
put far away from others. In the supergravity, the effect of the probe is described 
as a perturbation around black brane background, specified by the boundary condition $\{h_i\}$. 
In gauge theory side, it corresponds to $U(N+1)$ super Yang-Mills theory 
whose symmetry is broken to $U(N)\times U(1)$. The supergravity modes $\{h_i\}$ couple to 
operators ${\cal O}_i$ in the gauge theory. 
For $p=0$, precise dictionary between field theory operators and supergravity modes can be determined 
by looking at 
how ``Matrix theory currents" couple to bulk gravity modes \cite{Taylor}:   
one considers the D0-brane action in weakly curved background, which is the Born-Infeld action, 
expand it around the flat space to linear order in supergravity modes and see 
how they couple to Matrix theory operators.  
For example, bulk metric couples to 
the energy-momentum tensor.  
This observation leads to $p=0$ analogue of the GKPW proposal \cite{SY}
\begin{eqnarray}
e^{-S_{SG}[h]}
=
\left\langle
\exp\left(
\int dt\sum_i h_i(t){\cal O}_i(t)
\right)
\right\rangle_{YM},  
\end{eqnarray}
where $S_{SG}[h]$ is the supergravity action in black zero-brane background 
as a functional of boundary values of the fields $\{h_i\}$.  
The l.h.s. is evaluated by solving the classical equation of motion 
by imposing an appropriate boundary condition specified by $\{h_i\}$. 
In order for this approximation to be valid, 
distance between two operators must lie in an appropriate range, 
so that only weakly curved and small $g_s$ region of the bulk 
contribute to the correlation function. 
From this one can calculate correlation functions.  
At zero temperature, 
operators corresponding to supergravity modes are expected to follow a power law \cite{SY}
\begin{eqnarray}
\Big\langle
{\cal O}(t){\cal O}(t')
\Big\rangle
\propto
\frac{1}{|t-t'|^{2\nu+1}} \ ,
\label{coordspace} 
\end{eqnarray}
although the theory is not conformal\footnote{
The power law appears due to the generalized conformal symmetry,  
which is explained below. 
}.  
The region where the supergravity approximation is valid can be found as follows.  
As the separation between two operators becomes large, 
the supergravity mode propagates deeper into the bulk, so that 
it picks up the effect of large $g_s$ at small $U$. 
On the other hand, if separation is too small, 
contribution from large $U$ region becomes important 
because the radial coordinate is related to the energy scale in the gauge theory.  
From these considerations, it turns out the supergravity approximation is justified at 
$\lambda^{-1/3} \ll |t-t'|\ll
\lambda^{-1/3} N^{10/21}$ \cite{SY}. 
Note however that this is just a sufficient condition and 
the result of the approximation may extends beyond this region. 
As we will see numerically, this is the case indeed.

Another useful quantity which can be studied by using the duality is 
the supersymmetric Wilson loop \cite{Rey:1998ik}\cite{Witten:1998zw}. 
At finite temperature, the loop winding on temporal circle (Polyakov loop), 
\begin{eqnarray}
W
=
P\exp\left(
\int dt\left(
iA(t)+n_iX_i(t)
\right)
\right), 
\end{eqnarray} 
where $\vec{n}=(n_1,\cdots,n_9)$ is an arbitrary unit vector,  
is an order parameter of the confinement-deconfinement transition \cite{Witten:1998zw}. 
Scalar fields appears because strings pull D-branes.  
In gravity side, $\langle W\rangle$ is identified to the exponential of the area of 
the minimal string world-sheet ending on the loop \cite{Rey:1998ik}, 
\begin{eqnarray}
\log\langle W\rangle
=
1.89 \left(T\lambda^{-1/3}\right)^{-3/5}+\cdots, 
\end{eqnarray}
where $\cdots$ are possible logarithmic and constant corrections. 

In AdS/CFT correspondence, CFT operators are put on the `boundary' of the near horizon region, 
$U\to\infty$. This procedure seems to be subtle, however, because one has to take the near horizon 
limit in order to establish the correspondence and `boundary' is the place where this 
procedure may fail. It is widely believed that the conformal symmetry saves the situation; 
if the correspondence holds at small distance, it extends to longer distance 
as long as the conformal symmetry exists both in gauge theory and in the supergravity. 
However, the system is not conformal when $p\neq 3$. Then what can protect the correspondence?  
Actually the geometry (\ref{near extremal metric}) is invariant under 
the {\it generalized scale transformation}
\begin{eqnarray}
t,y_{\parallel}\ \  & \to &\ \  c^{-1}t, c^{-1}y_{\parallel},\\
U\ \  & \to &\ \  cU,\\
g_{YM}^2\ \  & \to &\ \  c^{3-p}g_{YM}^2, 
\end{eqnarray}
where $c>0$. There is a counterpart in gauge theory, 
\begin{eqnarray}
A_\mu,X_i\ \ & \to & \ \ cA_\mu,cX_i, \\
t,x\ \ & \to & \ \ c^{-1}t, c^{-1}x, \\
g_{YM}^2\ \  & \to &\ \  c^{3-p}g_{YM}^2. 
\end{eqnarray}
This generalized conformal symmetry serves as an alternative to the conformal symmetry 
\cite{Jevicki:1998qs,Jevicki,SY}\cite{Azeyanagi:2008mi}. 
Note that, for $p\neq 3$, gauge coupling is also scaled reflecting 
the fact that it is dimensionful. (The effective coupling $\lambda T^{-(3-p)}$ 
is invariant.) 
Therefore, in string theory, string coupling 
constant $g_s$ is rescaled. 
Hence it is not a ``symmetry" in the usual sense; 
it should be interpreted as a transformation in the whole moduli of string/M theory.

\section{Setup for numerical simulation}\label{sec:simulation_setup}

The action of the maximally supersymmetric matrix quantum mechanics  
is obtained formally by dimensionally
reducing 10d super Yang-Mills theory to 1d:
\begin{eqnarray}
S 
=
\frac{1}{g_{YM}^2} \int_0^{\beta}  
d t \, 
Tr 
\bigg\{ 
\frac{1}{2} (D_t X_i)^2 - 
\frac{1}{4} [X_i , X_j]^2  
+ \frac{1}{2} \psi_\alpha D_t \psi_\alpha
- \frac{1}{2} \psi_\alpha \gamma_i^{\alpha\beta} 
 [X_i , \psi_\beta ]
\bigg\} \ ,
\label{cQM}
\end{eqnarray}
where $D_t  = \partial_t
  - i \, [A(t), \ \cdot \ ]$ represents the covariant derivative
with the gauge field $A(t)$ being an $N\times N$ Hermitian matrix.
It can be viewed as 
a one-dimensional $U(N)$ gauge theory with adjoint matters.
The bosonic matrices $X_i(t)$  $(i=1,\cdots,9)$
come from spatial components of the 10d gauge field,
while the fermionic matrices $\psi_\alpha(t)$
$(\alpha=1,\cdots , 16)$ come from
a Majorana-Weyl spinor in 10d.
The $16\times 16$ matrices $\gamma_i$
in (\ref{cQM}) act on spinor indices and
satisfies the Euclidean Clifford algebra
$\{ \gamma_i,\gamma_j \}= 2\delta_{ij}$.
When we study the energy density and the Wilson loop, we are interested  
in the finite temperature. 
Therefore, we impose periodic and anti-periodic
boundary conditions on the bosons and fermions, respectively.
The extent $\beta$ in the Euclidean time 
direction then corresponds to the inverse
temperature $\beta \equiv 1/T$. 
For evaluation of correlators, we impose periodic boundary condition 
for both bosons and fermions. 
The coupling constant $g_{YM}$ in (\ref{cQM})
can always be scaled out by
an appropriate rescaling of the matrices and 
the time coordinate $t$.
We take $g_{YM} = \frac{1}{\sqrt{N}}$ $(\lambda=1)$ without loss 
of generality.

We take the static diagonal gauge
$A(t) = \frac{1}{\beta} {\rm diag}
(\alpha_1 , \cdots , \alpha_N)$,
where $\alpha_a$ can be chosen to 
satisfy
the constraint 
$\max_a (\alpha_a) - \min_a (\alpha_a) 
\le 2\pi$
by using the large gauge transformation
with a non-zero winding number.
We have to add to the action a term
\begin{eqnarray}
S_{\rm FP} =
- \sum_{a<b} 2 \ln 
\left| \sin \frac{\alpha_a - \alpha_b}{2}
\right|  \ , 
\end{eqnarray}
which appears from the Faddeev-Popov procedure,
and the integration measure for $\alpha_a$
is taken to be uniform.

At finite temperature, we make a Fourier expansion 
\begin{eqnarray}
X_i ^{ab} (t) = \sum_{n=-\Lambda}^{\Lambda} 
\tilde{X}_{i n}^{ab} e^{i \omega n t} \ ; \
\psi_\alpha ^{ab} (t) = \sum_{r=-(\Lambda-1/2)}^{\Lambda-1/2}
\tilde{\psi}_{\alpha r}^{ab} e^{i \omega r t} \ . 
\end{eqnarray} 
The indices $n$ and $r$ take integer and
half-integer values, respectively,
corresponding to the imposed 
boundary conditions.\footnote{ 
When we impose periodic boundary condition for fermions, 
they are expanded as 
\begin{eqnarray}
\psi_\alpha ^{ab} (t) = \sum_{r=-\Lambda}^{\Lambda}
\tilde{\psi}_{\alpha n}^{ab} e^{i \omega n t} \ .
\end{eqnarray} 
}
Introducing a shorthand
notation
\begin{eqnarray}
\Bigl(f^{(1)}  \cdots  f^{(p)}\Bigr)_n 
\equiv \sum_{k_1 + \cdots + k_{p}=n}
f^{(1)}_{k_1} \cdots f^{(p)}_{k_p} \ ,
\end{eqnarray}
we can write the action 
(\ref{cQM}) as
$S=S_{\rm b}+S_{\rm f}$, where
\begin{eqnarray}
S_{\rm b}
=  N \beta
\Bigg[
\frac{1}{2} \sum_{n=-\Lambda}^{\Lambda} 
\left( n \omega - \frac{\alpha_a - \alpha_b}{\beta} 
\right)
^2   \tilde{X}_{i , -n}^{ba} \tilde{X}_{i n}^{ab}
- \frac{1}{4} 
 Tr \Bigl( [ \tilde{X}_{i} , \tilde{X}_{j}]^2  \Bigr)_0
\Bigg] , 
\nonumber \\
S_{\rm f}
= \frac{1}{2}
N \beta \sum_{r=-(\Lambda-1/2)}^{\Lambda-1/2} \Biggl[
i 
\left(
r \omega - \frac{\alpha_a - \alpha_b}{\beta}  
\right)
\tilde{\bar{\psi}}_{\alpha r}^{ba} \tilde{\psi}_{\alpha r}^{ab} 
- (\gamma_i)_{\alpha\beta}
 Tr \Bigl\{ \tilde{\bar{\psi}}_{\alpha r} \Bigl(
[ \tilde{X}_{i},\tilde{\psi}_{\beta}] \Bigr)_r \Bigr\} \Biggr] \ .
\label{bfss_action_cutoff}
\end{eqnarray} 
The fermionic action $S_{\rm f}$ may be
written in the form 
$S_{\rm f}
= \frac{1}{2} {\cal M}_{A \alpha r ; B \beta s}
\tilde{\psi}_{\alpha r}^A \tilde{\psi}_{\beta s}^B
$,
where we have expanded
$\tilde{\psi}_{\alpha r}
= \sum_{A=1}^{N^2} \tilde{\psi}_{\alpha r}^A t^A$
in terms of $U(N)$ generators $t^A$. 
Integrating out the fermionic variables, one obtains
the Pfaffian ${\rm Pf}{\cal M}$, which is complex
in general. 
However, we observe that it is actually
real positive with high accuracy in the temperature
regime studied in \S~\ref{sec:energy}. 
At very low temperature relevant for \S~\ref{sec:wilson}, and 
with periodic boundary condition adopted in \S~\ref{sec:GKPW}, 
the phase of the Pfaffian does fluctuate violently. 
Here we simply neglect the phase and use 
$|{\rm Pf}{\cal M}|
= {\rm det} ( {\cal D}^{1/4})$, 
where ${\cal D}={\cal M}^\dag {\cal M}$, 
instead of ${\rm Pf}{\cal M}$. 
Surprisingly, it leads to the results expected by the duality\footnote{
The agreement suggests the phase quenching is valid, but is not a ``proof". 
To justify it one should look at the correlation between phase and values of 
physical observables \cite{HNSY_in_preparation}. 
}.

To simulate this system treating fermions fully dynamically, we adopt 
the RHMC algorithm \cite{Clark:2003na}. 
The trick of the RHMC is to
use the rational approximation
$x^{-1/4} \simeq
b_0 + \sum_{k=1}^{Q}
\frac{a_k}{x+b_k} 
$,
which has sufficiently small relative error 
within a certain range required by the system
to be simulated.
%
(The real positive parameters 
$a_k$ and $b_k$ can be
obtained by a code \cite{Clark-Kennedy} based on 
the Remez algorithm.)
Then the Pfaffian is replaced by
$
|{\rm Pf}{\cal M}|
= \int dF dF^* \exp\left(-S_{\rm PF}\right) 
$, where
\begin{eqnarray}
S_{\rm PF} =
b_0 F^* F + \sum_{k=1}^{Q}
a_k F^* ({\cal D}+b_k)^{-1} F \ ,
\label{PF-pf}
\end{eqnarray} 
using the auxiliary complex bosonic variables $F$,
which are called the 
pseudo-fermions.

We apply the usual
HMC algorithm to the whole system
as described in \cite{Catterall:2007fp},
except that now we introduce
the momentum variables conjugate to
the pseudo-fermions $F$ as well 
as the bosonic matrices $\tilde{X}_i$ 
and the gauge variables $\alpha_a$. 

When we solve the auxiliary classical
Hamiltonian dynamics, low momentum modes tend to evolve  
larger amount compared to high momentum modes. 
Therefore, by taking the step sizes of the evolution of 
lower momentum modes larger, configuration space can be swept out 
more efficiently. This method is called the Fourier acceleration 
\cite{Catterall:2001jg}. In the lattice gauge theory, 
in order to apply the Fourier acceleration one has to transform the 
configuration to the momentum representation. 
On the other hand, in the momentum cutoff method, 
because we are working directly in the momentum representation, 
the Fourier acceleration can be implemented without any additional cost. 
Thanks to this advantage the simulation cost can be reduced drastically.  

The main part of the computation
comes from  
solving a linear system 
$({\cal D}+b_k) \chi = F \quad
(k=1, \cdots , Q )
$. 
We solve the system
for the smallest $b_k$
using the conjugate gradient
method, which reduces the problem to
the iterative multiplications of ${\cal M}$
to a pseudo-fermion field,
each of which requires O($\Lambda^2 N^3$)
arithmetic operations if implemented carefully.
The solution for larger $b_k$'s 
can be obtained as by-products
using the idea of the multi-mass 
Krylov solver \cite{Jegerlehner:1996pm}. 
This avoids the factor
of $Q$ increase of the computational effort.

\section{Results}\label{sec:results}

\subsection{(In)stability of the black hole}\label{sec:stability}
Because the theory is supersymmetric, it has a flat direction along which 
scalar fields commute each other. As a result, partition function is divergent 
even at finite volume. 
Then how can we study this system numerically? 

String theory interpretation of the flat direction is obvious. 
Because scalar eigenvalues corresponds to the positions of D0-branes, 
it corresponds to a gas of D0-branes. Once they spread widely, 
there is no force between them because of the supersymmetry and hence 
each D0-brane propagates freely. Black 0-brane, which is considered in 
the gauge/gravity duality, is a bound state of many D0-branes. 
In the matrix model, it is a bound state of scalar eigenvalues. 
Because the supergravity approximation is valid at large-$N$, and there 
the black 0-brane is stable, we expect there is a stable bound state of 
eigenvalues at large-$N$. At finite-$N$, because the string coupling 
constant is finite, closed string can propagate and sometimes it escapes 
from the black 0-brane. In other words, scalar eigenvalues can run away 
from the bound state. Hence the bound state should be at most metastable at finite-$N$, 
i.e. black hole is unstable due to the quantum stringy effect.

This is exactly what we observe in numerical simulation. 
For fixed $N$, we observe a metastable state at high and low temperature, 
while at intermediate temperature eigenvalues spreads quickly. 
To obtain expectation values, we use only configurations in metastable states. 
As $N$ increases, the metastable state becomes stabler and 
it exists in wider parametric region. 
 
\subsection{Energy vs ADM mass}\label{sec:energy}
In order to calculate the energy density efficiently, we use 
a trick introduced in \cite{Catterall:2007fp}. 
In the momentum cutoff prescription, one obtains \cite{Anagnostopoulos:2007fw} 
\begin{eqnarray}
E
=
-3T\left\{
\langle S_b\rangle
-
\frac{9}{2}\left(
(2\Lambda+1)N^2-1
\right)
\right\}. 
\end{eqnarray}

As we will see shortly, our data is precise enough so that 
we can evaluate the deviation from the supergravity limit quantitatively. 
In the dual gravity point of view, the deviation at large-$N$ and 
at finite 't Hooft coupling corresponds to the string $\alpha'$ correction. 
Higher derivative corrections to the type IIA supergravity 
start with $\alpha'^3$ order \cite{GW}. By a simple dimensional counting, 
it takes the form $(\alpha'/R^2)^3$, where $R$ is the curvature radius 
of the black brane background, which translates to $(T\lambda^{-1/3})^{9/5}$. 
Therefore, at large-$N$, correction to 
the supergravity limit $E/N^2=7.41(T\lambda^{-1/3})^{14/5}$ should be 
$c(T\lambda^{-1/3})^{23/5}$, where $c$ is an unknown constant. 

\begin{figure}[htb]
\begin{center} 
\includegraphics[height=6cm]{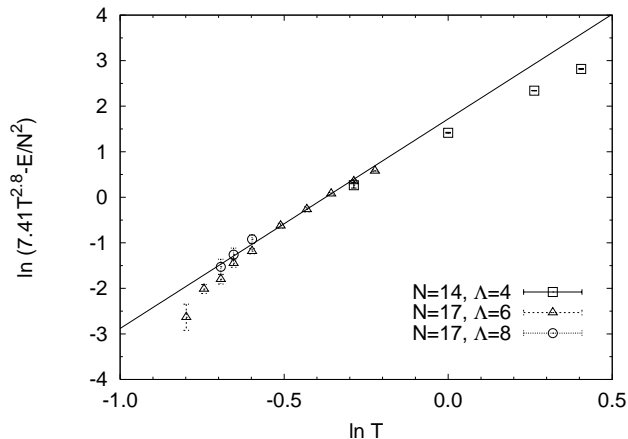} 
\end{center}
\caption{
The deviation of the internal energy
$\frac{1}{N^2} E$ from the leading term
$7.41 \, T^{\frac{14}{5}}$
is plotted against the temperature
in the log-log scale for $\lambda=1$.
The solid line represents 
a fit to a straight line with the slope $23/5$
predicted from the $\alpha '$ corrections 
on the gravity side.
}
\label{energy}
\end{figure}

\begin{figure}[htb]
\begin{center}
\includegraphics[height=6cm]{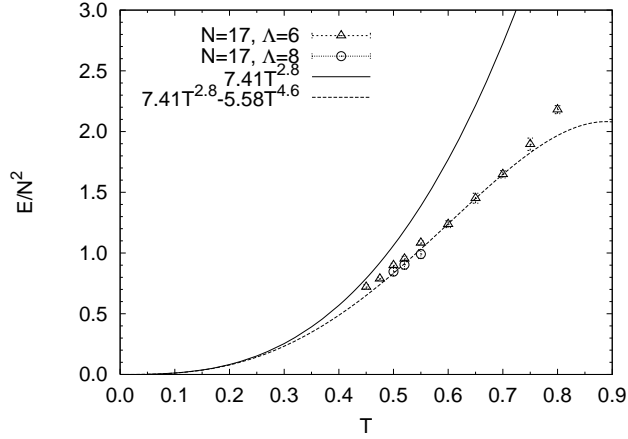}
\end{center}
\caption{
The internal energy $\frac{1}{N^2} E$
is plotted against $T$ for $\lambda=1$.
The solid line represents
the leading asymptotic behavior at small $T$
predicted by the gauge-gravity duality.
The dashed line represents
a fit to the behavior (\ref{EvsTsub})
including the sub-leading term
with $p=23/5$ (fixed) and $C=5.58$.
}
\label{energy2}
\end{figure}

In Fig.~\ref{energy} and Fig.~\ref{energy2} we plot the energy density 
obtained from our simulation \cite{Hanada:2008ez}. 
Fig.~\ref{energy} is a log-log plot of $7.41T^{14/5}-E/N^2$ versus $T$. 
At sufficiently low temperature, a power law behavior can be seen 
clearly (i.e. data points form a straight line), which indicates that 
higher order corrections are negligible. (A deviation from a power law 
at very low temperature is merely a finite cutoff effect. Indeed 
by increasing the momentum cutoff from $\Lambda=6$ to $\Lambda=8$ 
data points go closer to the straight line.) 
By fitting the result with an ansatz 
\begin{eqnarray}
\frac{E}{N^2}
=
7.41T^{14/5}
-
cT^p, 
\label{EvsTsub}
\end{eqnarray}
we obtain $p=4.58(3)$ and $c=5.55(7)$. If we fix $p=23/5$ instead, we obtain 
$c=5.58(1)$. It strongly suggests that the duality is correct including 
the $\alpha'$ correction.  
Also, by assuming the duality is correct at this level, Monte-Carlo simulation 
provides very powerful tool to study the stringy correction to the black hole 
thermodynamics. Note that the coefficient $c$ cannot be determined from gravity side,  
because the detail of the higher derivative correction to 
the type IIA supergravity is not known. Therefore our numerical result can be 
regarded as a {\it prediction from gauge theory side}. Usually people use the gauge/gravity 
duality to calculate something difficult (strongly coupled gauge theory) 
by dealing with easier problem (supergravity). However here we can use something difficult  
to study something more difficult (string theory).

\subsection{Wilson loop}\label{sec:wilson}
Next let us consider the Wilson loop \cite{Hanada:2008gy}. 
It turned out that agreement with dual gravity prediction emerges 
only at much lower temperature compared to the case of the energy density. 
In such a low temperature region, because we have to take the cutoff $\Lambda$ 
to be large, we could study only rather small values of $N$, up to $N=8$. 

\begin{figure}[htb]
\begin{center}
\includegraphics[height=6cm]{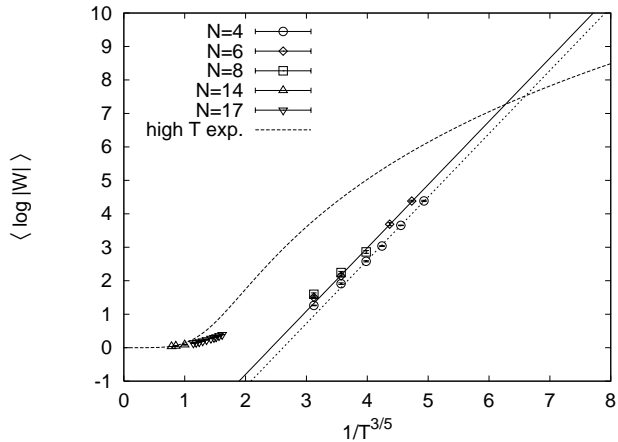}
\end{center}
\caption{
The plot of
$\langle\log|W|\rangle$ for $\lambda=1$
against $T^{-3/5}$. 
The cutoff $\Lambda$ is chosen as follows: 
$\Lambda=12$ for $N=4$; $\Lambda=0.6/T$ for $N=6,8$; 
$\Lambda=4$ for $N=14$; $\Lambda=6$ for $N=17$.
The dashed line represents the results
of the high-temperature expansion
up to the next-leading order
with extrapolations to
$N=\infty$, which are obtained by applying the method 
in \cite{HTE}.
The solid line and the dotted line represent fits
for $N=6$ and $N=4$ respectively,
to straight lines with the slope 1.89 
predicted from the gravity side at the leading order.
}
\label{fig:logW}
\end{figure}

In Fig.~\ref{fig:logW}, data points are shown in high and low temperature regions. 
At intermediate temperature, simulation with modest values of $N$ is unstable 
because of the scalar instability. Note that we have evaluated $\langle \log|W|\rangle$ 
instead of $\log\langle W\rangle$ in order to reduce numerical error. 
We have taken the absolute value because at finite-$N$ arbitrary phase factor can emerge 
due to the center symmetry and hence $\langle W\rangle =0$. At large-$N$, 
tunneling to different phase is suppressed and $\langle W\rangle$ agrees with $\langle |W|\rangle$ 
up to a fixed phase factor. That we have taken a logarithm before taking the expectation value 
can be justified at large-$N$, where fluctuation is suppressed.

As can be seen from Fig.~\ref{fig:logW}, logarithm of the Wilson loop behaves as  
$1.89 T^{-3/5} + const$. In fact it is difficult to distinguish 
constant and $\log T$ from our data; from stringy correction and quantum fluctuation 
of the string world-sheet, both constant and logarithmic corrections should arise. 
We consider the ``constant" is actually sum of these corrections. 

\subsection{Correlators}\label{sec:GKPW}
Next we show simulation result for two-point functions \cite{Hanada:2009ne}. 
Here we consider a class of operators $J^{+ij}$ given by 
\begin{eqnarray}
J^{+ij}_{l, i_1\cdots i_l}
\equiv
\frac{1}{N}Str\left(F_{ij}X_{i_1}\cdots X_{i_l}\right)
\qquad
(l\ge 1),    
\end{eqnarray}
which couples to NS-NS 2-form and R-R 1-form. 
The exponent $\nu$ predicted from the supergravity is \cite{SY}
\begin{eqnarray}
\nu=\frac{2l}{5}. 
\end{eqnarray} 
 
In the simulation, first we obtain the correlation function in the momentum space, 
\begin{eqnarray}
\left\langle
\tilde{J}_l^+(p)
\tilde{J}_l^+(-p)
\right\rangle
\qquad(p=2\pi n/\beta; \ n=0,1,2,\cdots).  
\end{eqnarray}
Because of the existence of the momentum cutoff $\Lambda$, the result can be trusted 
only at small $p$. In order to estimate the correlators at large $p$,  
we assume the behavior 
\begin{eqnarray}
\left\langle\tilde{J}_l^+(p)\tilde{J}_l^+(-p)\right\rangle
=
\frac{a}{p^2}, 
\label{ansaz:extrapolation in momentum space}
\end{eqnarray}
where $a$ is a constant, 
and we determine coefficients $a$ by fitting a few points close to $\Lambda$.   

In order to obtain the correlators in the coordinate space, 
we perform the Fourier transformation, 
\begin{eqnarray}
\langle J_l^+(t)J_l^+(0)\rangle
=
\left\langle\tilde{J}_l^+(0)\tilde{J}_l^+(0)\right\rangle
+
\sum_{p>0}
2\cos(pt)
\left\langle\tilde{J}_l^+(p)\tilde{J}_l^+(-p)\right\rangle. 
\end{eqnarray}
We terminated the sum w.r.t. $p$ at $p=\frac{2\pi}{\beta}\times 1000$. 

\begin{figure}[htbp]
\begin{center}
\scalebox{0.4}{
\rotatebox{-90}{
\includegraphics{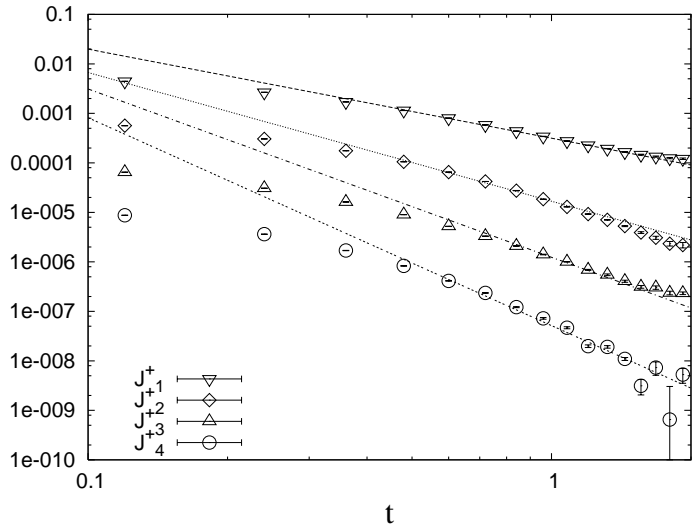}}}
\caption{
Log-log plot of the correlator $\langle J^+_i(t)J^+_i(0)\rangle$ $(i=1,2,3,4)$ 
for $N=3, \Lambda=16, \beta=4$. 
}\label{fig:Jp_N3C16T025} 
\end{center} 
\end{figure}

As a concrete example, we consider correlators at 
$N=3, \Lambda=16, \beta=4$. For $n>12$, we obtain the two-point function by 
extrapolating the data with the ansatz 
(\ref{ansaz:extrapolation in momentum space}). 
For the fitting, $n=10,11$ and $12$ are used.    
We evaluated the error by using the Jack-knife method, 
by dividing samples to four bins.  
The result after the Fourier transformation is shown in Fig.~\ref{fig:Jp_N3C16T025}. 
Straight lines represents a power law behavior proportional to $1/t^{2\nu+1}$, 
where $\nu=2l/5$ is the prediction from the supergravity. 
Surprisingly, the expected power appears at such a small value of $N$. 
In fact 
the power law is reproduced with remarkable precision, already at $N=2$ \cite{HNSY_in_preparation}.  
Furthermore, the power law extends beyond the parameter region discussed in \S~\ref{sec:duality}. 
It is very interesting because it suggests the gravity prediction can shed a light 
on deep IR region, which is relevant for the matrix theory conjecture for M-theory \cite{BFSS}. 

\section{Conclusion and discussions}
In this talk we explained recent Monte Carlo simulations for 
maximally supersymmetric matrix quantum mechanics, in the context of the superstring theory\footnote{
The same system has been studied in \cite{Catterall-Wiseman} and qualitatively consistent 
results have been obtained. 
}. 
The data reproduces predictions by the gauge/gravity duality precisely, 
and furthermore it provides a nontrivial prediction for stringy correction 
to black hole thermodynamics. 

There are many future directions. First of all, the same simulation techniques 
can be applied to other quantum mechanical system as well. 
It would be nice if one can learn nonperturbative aspects of interesting theories 
which cannot be calculated analytically. It is also interesting to study 
theories in higher than one dimension. For example three-dimensional maximally supersymmetric 
Yang-Mills theory is realized \cite{Maldacena:2002rb} around fuzzy sphere background of 
the plane wave matrix model \cite{Berenstein:2002jq}, and hence can be studied 
with the momentum cutoff method\footnote{
A lattice formulation can be found in \cite{KU_16SUSY}.}. 
Another background in the same matrix model is argued to correspond to 4d ${\cal N}=4$ SYM 
in the planar limit \cite{Ishii:2008ib,Nishimura:2009xm}. In two dimensions, lattice formulation works 
without fine tuning \cite{KU_16SUSY,2d_16SUSY,Suzuki:2005dx} to all order in perturbation theory, and 
numerical studies so far confirms it works also at nonperturbative level. 
(For 4-SUSY system, absence of fine tuning is confirmed in \cite{Hanada:2009hq}, 
by using dynamical fermion, for two independent lattice formulations. 
The conservation of the supercurrent has also been confirmed \cite{Kanamori:2008bk}.)  
From string theory point of view, 
2d SYM is related \cite{Susskind:1997dr,Aharony:2004ig} to the black hole/black string transition   
\cite{Gregory:1993vy}. Recent numerical study in this context can be found in \cite{Hanada:2009hq,CJW}. 
Finally, 4d ${\cal N}=4$ SYM at finite-$N$ level is realized \cite{Hanada:2010kt} 
by combining a matrix model technique \cite{Maldacena:2002rb} and two-dimensional lattice\footnote{
In usual lattice formulation, there is no obvious reason for the absence of the fine tuning. 
However explicit calculation shows it is absent at one-loop level, and an attempt to provide  
all-loop argument is going on \cite{Catterall:2010jh}. }. 
Together with \cite{Ishii:2008ib}, it provides nonperturbative tools to study 
$AdS_5/CFT_4$ correspondence. Monte-Carlo study of these models is of crucial importance\footnote{
For another approach to $AdS_5/CFT_4$ using simplified model see section V of \cite{Berenstein:2010xw} 
and references therein. 
}. 

\section*{Acknowledgments}
I would like to thank all the collaborators involved in the projects 
reported here. 
In particular, I am grateful to Jun Nishimura for useful comments on this proceedings.  
I would also like to thank the organizers  
of the Workshop ``Supersymmetric Quantum Mechanics and Spectral Design" 
for warm hospitality.



\end{document}